\documentclass[12pt]{article}

\newcommand{\bce}{\begin{center}}
\newcommand{\ece}{\end{center}}
\newcommand{\beq}{\begin{equation}}
\newcommand{\eeq}{\end{equation}}
\newcommand{\bea}{\vspace{0.25cm}\begin{eqnarray}}
\newcommand{\eea}{\end{eqnarray}}

\newcommand{\ba}{\begin{array}}
\newcommand{\ea}{\end{array}}

\def\lsim{\mathrel{\rlap{\lower4pt\hbox{\hskip1pt$\sim$}}
    \raise1pt\hbox{$<$}}}         
\def\gsim{\mathrel{\rlap{\lower4pt\hbox{\hskip1pt$\sim$}}
    \raise1pt\hbox{$>$}}}         

\def\Pom{{\bf I\!P}}

\def\lsim{\mathrel{\rlap{\lower4pt\hbox{\hskip1pt$\sim$}}
    \raise1pt\hbox{$<$}}}         
\def\gsim{\mathrel{\rlap{\lower4pt\hbox{\hskip1pt$\sim$}}
    \raise1pt\hbox{$>$}}}         

\def\Pom{{\bf I\!P}}

 \advance\oddsidemargin  by -0.9in
  \advance\evensidemargin by -0.9in
  \advance\topmargin      by -0.20in
\makeindex
\begin{document}

\phantom{.}\hspace{10.5cm}{\large \bf 24 March 2000}
\vspace{1.5cm}\\
\begin{center}
{\Large \bf
Anomalous $t$-dependence
in diffractive electroproduction
of 2S radially excited
light vector mesons at HERA} \\
\vspace*{1.5cm} 
{\large \bf J.~Nemchik}
\vspace*{0.5cm} \\ 
{\it Institute of Experimental Physics, Slovak Academy of Sciences, \\
Watsonova 47, 04353 Ko\v sice, Slovakia} \\
\vspace*{3.5cm} 
\begin{minipage}[h]{13cm}
\centerline{\Large \bf 
Abstract }
\vspace*{1.5cm}

%
Within the color dipole gBFKL dynamics applied
to the diffraction slope,
we predict an 
anomalous $t$ dependence of the
differential cross section as a function of energy and $Q^{2}$
for production of radially excited $V'(2S)$ light
vector mesons in contradiction with
a well known standard monotonous
$t$- behaviour for $V(1S)$ mesons.
The origin of this phenomenon is based on the interplay of 
the nodal structure of 
$V'(2S)$ radial wave function with
the energy and dipole size dependence 
of the color dipole cross
section and of the
diffraction slope.
We present how a different position of the node in $V'(2S)$ wave function
leads to a different form of anomalous $t$- behaviour  
of the differential cross section and 
discuss a possibility how to determine this
position from the low energy and HERA data.
\end{minipage}
\end{center}
\pagebreak
\setlength{\baselineskip}{0.55cm}
%
%

The main goal of this paper is a demonstration
of further salient features of the node effect 
\cite{KZ91,KNNZ93,NNZanom,NNPZ97,NNPZZ98}
coming from the nodal structure of
radial wave function for $V'(2S)$ vector mesons 
in conjunction with the gBFKL phenomenology of the
diffraction slope 
\cite{NZZslope,NZZspectrum,NNPZZ98}
leading to
anomalous $t$ dependence of the differential
cross section for $V'(2S)$ production in contrast
with the standard monotonous $t$ behaviour 
of $d\sigma(\gamma^{*}\rightarrow V)/dt$ for $V(1S)$
production.

Diffractive photo- and electroproduction of 
ground state $V(1S)$ and radially excited $V'(2S)$
vector mesons,
%
%
\beq
\gamma^{*}p \rightarrow V(V')p ~~~~~~~~V = \rho, \Phi, \omega, J/\Psi, \Upsilon
... ~~~~~~~~(V' = \rho', \Phi', \omega', \Psi', \Upsilon'...)\, ,
\eeq
%
%
at high c.m.s. energy $W=\sqrt{s}$
intensively studied by the recent experiments
at HERA represents one of
a main source for a further development of the pomeron physics.
The pomeron exchange 
in diffractive leptoproduction of vector mesons
at high energies
has been intensively studied
\cite{DL,KZ91,Ryskin,KNNZ93,KNNZ94,NNZscan,Brodsky,Forshaw,GLM} 
within the framework of perturbative QCD (pQCD).

The standard approach to the pQCD is based on the BFKL equation
\cite{Kuraev,Balitsky,Lipatov}, which represents the integral
equation for the leading-log$s$ (LLs) evolution of the
gluon distribution, formulated
in the scaling approximation of the infinite gluon correlation
radius, $R_{c}\rightarrow\infty$, (massless gluons) and
of the fixed running coupling,
$\alpha_{S}=const$.
Later, however, a novel $s$-channel approach
to the $LLs$ BFKL equation (running gBFKL approach)
has been developed \cite{NZ94,NZZ94} in terms of the color
dipole cross section
$\sigma(\xi,r)$ (hereafter $r$ is the transverse size of
the color dipole, $\xi = log ({{W^{2}+Q^{2}} \over {m_{V}^{2}+Q^{2}}})$
is the rapidity variable) and incorporates consistently
the asymptotic freedom (AF) (i.e. the running QCD coupling $\alpha_{S}(r)$)
and the finite propagation radius $R_{c}$ of perturbative gluons.
The freezing of $\alpha_{S}(r)$, $\alpha_{S}(r)\le \alpha_{S}^{fr}$,
and the gluon correlation radius $R_{c}$ represents the
nonperturbative parameters, which describe the transition from
the soft (nonperturbative, infrared) to the hard
(perturbative) region.

The details of the gBFKL phenomenology of diffractive
electroproduction of light vector mesons are presented
in the paper \cite{NNPZ97}. The color dipole phenomenology
of the diffraction slope for photo- and electroproduction
of heavy vector mesons has been developed in the paper \cite{NNPZZ98}.
The analysis of the diffractive production of light
\cite{NNZscan,NNPZ97} and heavy 
\cite{NNPZZ98} vector mesons at $t=0$ within
the gBFKL phenomenology 
shows that the
$1S$ vector meson production amplitude probes the color dipole cross
section at the dipole size $r\sim r_{S}$
({\it scanning phenomenon} \cite{NNN92,KNNZ93,KNNZ94,NNZscan}), 
where the scanning radius can be expressed through the scale
parameter $A$
%
\beq
r_{S} \approx {A \over \sqrt{m_{V}^{2}+Q^{2}}}\, ,
\label{eq:1}
\eeq
%
where $Q^{2}$ is the photon virtuality,
$m_{V}$ is the vector meson mass and $A\approx 6$.
Thus,
changing $Q^{2}$ and the mass of the produced
vector meson, one can
probe the dipole cross section $\sigma(\xi,r)$,
and the dipole diffraction slope $B(\xi,r)$,
and measure so the
effective intercept $\Delta_{eff}(\xi,r)
={\partial\log\sigma(\xi,r)/ \partial\xi}$ and
the local Regge slope
$\alpha_{eff}'(\xi,r)
={1 \over 2}{\partial B(\xi,r)/ \partial\xi}$
in a very broad range of the dipole
sizes, $r$.
This fact allows also to study the transition
between the perturbative (hard)
and nonperturbative (soft) regimes.

The experimental investigation 
of the electroproduction
of the radially excited ($2S$) vector mesons can extend
an additional 
information on the dipole cross section and on the dipole
diffraction slope.
The presence of the node in the
$2S$ radial
wave function leads to
a strong cancellation
of the dipole size contributions to the production amplitude
from the region above and below the node position,
$r_{n}$, in the $2S$ radial wave function \cite{KZ91,NNN92,NNZanom,NNPZ97}.
For this reason, the amplitudes of the
electroproduction of the $1S$ and $2S$ vector mesons probe $\sigma(\xi,r)$
and $B(\xi,r)$ in a different way.
The onset of strong node effect has been
demonstrated in Ref.~\cite{NNPZ97} in
electroproduction of radially excited 
light vector mesons leading 
to an anomalous $Q^{2}$ and energy dependence of the production
cross section.
The node effect is much weaker for the electroproduction of $2S$ heavy
vector mesons but still leads to a slightly different $Q^{2}$ and
energy dependence of the production cross section
for $\Psi'$ vs. $J/\Psi$ 
and to a nonmonotonic $Q^{2}$ dependence of
the diffraction slope at small
$Q^{2}\lsim 5$\,GeV$^{2}$ for $\Psi'$ production
\cite{NNPZZ98}.
Only for $\Upsilon'$ production, the node
effect is negligible small and gives approximately the
same $Q^{2}$ and energy behaviour of the production cross section
and practically the same diffraction slope at $t=0$ for
$\Upsilon$ and $\Upsilon'$ production \cite{NNPZZ98}.
Therefore, it is very important to explore farther the salient features
of the node effect with conjunction with the emerging gBFKL
phenomenology of the diffraction slope especially in production of
$V'(2S)$ light vector mesons where the node effect is
expected to be very strong.

There are two main reasons which affect the cancellation pattern
in the diffraction slope for $2S$ state.
The first reason is connected with the $Q^{2}$ behaviour
of the scanning radius $r_{S}$ (see (\ref{eq:1}));
for the electroproduction of $V'(2S)$ light vector mesons
at moderate $Q^{2}$ when the scanning radius
$r_{S}$ is close to $r_{n}$,
due to $\sim r^{2}$ behaviour of $B(\xi,r)$ \cite{NZZslope},
even a slight variation of
$r_{S}$ with $Q^{2}$ strongly changes the cancellation
pattern  and leads to an anomalous
$Q^{2}$ dependence 
of the forward diffraction slope, $B(t=0)$
\cite{NNPZZ98}.
The second reason is due to
different energy dependence
of $\sigma(\xi,r)$
at different dipole sizes $r$ coming from 
the gBFKL dynamics 
leading also to an
anomalous energy dependence of $B(t=0)$ for the $V'(2S)$ production.
This nonmonotonous energy and $Q^{2}$ dependence of the
diffraction slope 
for production of light vector mesons
will be detaily studied elsewhere \cite{prepare99}.

The effects mentioned above are sensitive to the
form of the dipole cross section.
In Ref.~\cite{NNPZdipole} we presented the first direct determination
of the color dipole cross section from the data on
the photo- and electroproduction of $V(1S)$ vector mesons.
So extracted dipole cross section is in a good agreement with
the dipole cross section obtained from gBFKL analysis
\cite{NZHera,NNZscan}.
This fact confirms a very reasonable choice
of the nonperturbative component of the dipole cross section
corresponding to a soft nonperturbative mechanism contribution
to the scattering amplitude.

In the present paper
we concentrate on the production of $2S$ radially
excited light vector mesons, where the node in the
radial wave function in conjunction with the
subasymptotic energy dependence of $B(\xi,r)$
leads to a
strikingly different 
$t$ dependence of the differential cross section
at different energies and $Q^{2}$
for the production of $V'(2S)$ vs.
$V(1S)$ vector mesons.
As was mentioned above,
due to a large value of the scale parameter in (\ref{eq:1}),
the large-distance
contributions to the production amplitude 
from the semiperturbative and nonperturbative
region of color dipoles, $r\gsim R_{c}$, becomes substantial
especially for light vector mesons.
Only the virtual $\rho^{0}$ and $\phi^{0}$ photoproduction
at $Q^{2}\gsim 100$\,GeV$^{2}$ can be treated as a purely
perturbative process, when the production
amplitude is dominantly contributed from the perturbative
region, $r\lsim R_{c}$.

Thus, in this paper 
we present the $Q^{2}$ and energy dependence
of the $t$- behaviour of the differential cross section
for electroproduction of the
ground state and radially excited (2S) light vector meson. 
and study how 
the position of the node in the radial wave function 
for (2S) vector mesons can be
extracted from the data. 
We present an exact prescription how the experimental
measurement of the 
$t$ dependent differential cross section for
$V'(2S)$ production could
distinguish between the undercompensation
and overcompensation scenarios of the $2S$ production amplitude
(see below).
The explicit form of that $t$- behaviour is connected
with the position of the node in radial wave function
for $V'(2S)$ vector mesons.

In the mixed $({\bf{r}},z)$ representation,
the high energy meson is considered as
a system of color dipole described by
the distribution
of the transverse separation ${\bf{r}}$ of the quark and
antiquark given by the $q\bar{q}$ wave function,
$\Psi({\bf{r}},z)$, where $z$ is
the fraction of meson's lightcone momentum
carried by a quark.
The Fock state expansion for the
relativistic meson starts
with the $q\bar{q}$ state and
the higher Fock states $q\bar{q}g...$
become very important at high energy $\nu$.
The interaction of the relativistic
color dipole of the dipole moment, ${\bf{r}}$, with the
target nucleon is quantified by the energy dependent color
dipole cross section, $\sigma(\xi,r)$,
satisfying
the gBFKL equation
\cite{NZ94,NZZ94} for the energy evolution.
This reflects the fact that 
in the leading-log ${1\over x}$ approximation the
effect of higher Fock states can be
reabsorbed into the energy dependence
of $\sigma(\xi,r)$.
The dipole cross section is flavor
independent and represents the universal
function of $r$ which describes
various diffractive processes in unified form.
At high energy, when the transverse separation, ${\bf{r}}$,
of the quark and antiquark is frozen during the interaction
process, 
the scattering
matrix describing the $q\bar{q}$-nucleon interaction
becomes diagonal
in the mixed $({\bf{r}},z)$-representation ($z$ is known also as
the Sudakov light cone variable). 
This diagonalization property is held even 
when the dipole size, ${\bf{r}}$, is large,
i.e. beyond the perturbative region of short distances.
The detailed discussion about the space-time
pattern of diffractive electroproduction of vector mesons
is presented in \cite{NNPZZ98,NNPZ97}.

Following an advantage of
the $({\bf{r}},z)$-diagonalization of the
$q\bar{q}-N$ scattering matrix, the
imaginary part of the production
amplitude for the real (virtual) photoproduction
of vector mesons
with the momentum transfer ${\bf{q}}$ can be represented in the 
factorized form
\beq
{\rm Im}{\cal M}(\gamma^{*}\rightarrow V,\xi,Q^{2},{\bf{q}})=
\langle V |\sigma(\xi,r,z,{\bf{q}})|\gamma^{*}\rangle=
\int\limits_{0}^{1} dz\int d^{2}{\bf{r}}\sigma(\xi,r,z,{\bf{q}})
\Psi_{V}^{*}({\bf{r}},z)\Psi_{\gamma^{*}}({\bf{r}},z)\,
\label{eq:2}
\eeq
whose normalization is
$
\left.{d\sigma/ dt}\right|_{t=0}={|{\cal M}|^{2}/ 16\pi}.
$
In Eq.~(\ref{eq:2}), 
$\Psi_{\gamma^{*}}({\bf{r}},z)$ and
$\Psi_{V}({\bf{r}},z)$ represent the
probability amplitudes
to find the color dipole of size, $r$,
in the photon and quarkonium (vector meson), respectively.
The color dipole distribution in (virtual) photons was
derived in \cite{NZ91,NZ94}.
$\sigma(\xi,r,z,{\bf{q}})$ 
is the scattering matrix for $q\bar{q}-N$ interaction and
represents the above mentioned color dipole cross section
for ${\bf{q}}=0$.
The color dipole cross section 
$\sigma(\xi,r)$ depends only on the dipole size $r$.
For small ${\bf{q}}$ considered in this paper,
one can safely neglect
the $z$-dependence of $\sigma(\xi,r,z,{\bf{q}})$ 
for light and heavy vector meson production
and set $z=\frac{1}{2}$.
This follows partially from the analysis within double gluon 
exchange approximation
\cite{NZ91} leading to a slow $z$ dependence of 
the dipole cross section.

The energy dependence of the dipole cross section is quantified
in terms of the dimensionless
rapidity, $\xi=\log{1\over x_{eff}}$, and $x_{eff}$ is
the effective value of the Bjorken variable 
%
%
\beq
x_{eff} =
\frac {Q^{2}+m_{V}^{2}}{Q^{2}+W^{2}} \approx
 \frac{m_{V}^{2}+Q^{2}}{2\nu m_{p}}\, ,
\label{eq:3}
\eeq
%
%
where $m_{p}$ and $m_{V}$ is the proton mass and mass of
vector meson, respectively.
Hereafter, we will write the energy dependence of the dipole
cross section in both variables,
either in $\xi$ or in $x_{eff}$ whenever convenient.

The production amplitudes for the
transversely (T) and the longitudinally (L) polarized vector mesons
with the small momentum transfer, $\bf{q}$,
can be written in more explicit form \cite{NNZscan,NNPZZ98}
\bea
{\rm Im}{\cal M}_{T}(x_{eff},Q^{2},{\bf{q}})=
{N_{c}C_{V}\sqrt{4\pi\alpha_{em}} \over (2\pi)^{2}}
\cdot~~~~~~~~~~~~~~~~~~~~~~~~~~~~~~~~~
\nonumber \\
\cdot \int d^{2}{\bf{r}} \sigma(x_{eff},r,{\bf{q}})
\int_{0}^{1}{dz \over z(1-z)}\left\{
m_{q}^{2}
K_{0}(\varepsilon r)
\phi(r,z)-
[z^{2}+(1-z)^{2}]\varepsilon K_{1}(\varepsilon r)\partial_{r}
\phi(r,z)\right\}\nonumber \\
 =
{1 \over (m_{V}^{2}+Q^{2})^{2}}
\int {dr^{2} \over r^{2}} {\sigma(x_{eff},r,{\bf{q}}) \over r^{2}}
W_{T}(Q^{2},r^{2})
\label{eq:4}
\eea
\bea
{\rm Im}{\cal M}_{L}(x_{eff},Q^{2},{\bf{q}})=
{N_{c}C_{V}\sqrt{4\pi\alpha_{em}} \over (2\pi)^{2}}
{2\sqrt{Q^{2}} \over m_{V}}
\cdot~~~~~~~~~~~~~~~~~~~~~~~~~~~~~~~~~
 \nonumber \\
\cdot \int d^{2}{\bf{r}} \sigma(x_{eff},r,{\bf{q}})
\int_{0}^{1}dz \left\{
[m_{q}^{2}+z(1-z)m_{V}^{2}]
K_{0}(\varepsilon r)
\phi(r,z)-
\partial_{r}^{2}
\phi(r,z)\right\} \nonumber \\
 =
{1 \over (m_{V}^{2}+Q^{2})^{2}}
{2\sqrt{Q^{2}} \over m_{V}}
\int {dr^{2} \over r^{2}} {\sigma(x_{eff},r,{\bf{q}}) \over r^{2}}
W_{L}(Q^{2},r^{2})
\label{eq:5}
\eea
where
\beq
\varepsilon^{2} = m_{q}^{2}+z(1-z)Q^{2}\,,
\label{eq:6}
\eeq
$\alpha_{em}$ is the fine structure 
constant, $N_{c}=3$ is the number of colors,
$C_{V}={1\over \sqrt{2}},\,{1\over 3\sqrt{2}},\,{1\over 3},\,
{2\over 3},\,{1\over 3}~~$ for 
$\rho^{0},\,\omega^{0},\,\phi^{0},\, J/\Psi, \Upsilon$ production,
respectively and
$K_{0,1}(x)$ are the modified Bessel functions.
The detailed discussion and parameterization 
of the lightcone radial wave function $\phi(r,z)$
of the $q\bar{q}$ Fock state of the vector meson
is given in \cite{NNPZ97}.

The terms $\propto \epsilon K_{1}(\epsilon r)\partial_{r}\phi({\bf r},z)$
for $T$ polarization  
and $\propto K_{0}(\epsilon r)\partial_{r}^{2}\Phi({\bf r},z)$
for $L$ polarization 
in the integrands of
(\ref{eq:4}) and (\ref{eq:5}) represent
the relativistic corrections 
which become important
at large $Q^{2}$ and for
the production of light vector mesons.
For the production of heavy quarkonia,
the nonrelativistic approximation can be used
with a rather high accuracy \cite{KZ91}.

For small dipole size and ${\bf{q}}=0$,
in the leading-log ${1\over x}$ approximation,
the dipole cross section can be related to the
gluon structure function $G(x,q^2)$ of the target nucleon
through
\beq
\sigma(x,r) =
\frac{\pi^2}{3}r^2\alpha_s(r)G(x,q^2) \, ,
\label{eq:7}
\eeq
where the gluon structure function enters at
the factorization scale,
$q^2 \sim {B\over r^2}$ \cite{Barone} with the parameter
$B\sim 10$ \cite{NZglue}.

The weight functions,
$W_{T}(Q^{2},r^{2})$ and
$W_{L}(Q^{2},r^{2})$, introduced in (\ref{eq:4}) and (\ref{eq:5})
have a smooth $Q^{2}$ behaviour \cite{NNZscan} and are very
convenient for the analysis of the scanning phenomenon.
They are
sharply peaked at $r\approx A_{T,L}/\sqrt{Q^{2}+m_{V}^{2}}$.
At small $Q^{2}$ the values of the scale parameter $A_{T,L}$ are 
close to $A \sim 6$, which follows from $r_{S}=3/\varepsilon$ with
the nonrelativistic choice $z=\frac{1}{2}$. 
In general, $A_{T,L} \geq 6$
and increases slowly with $Q^2$ \cite{NNZscan}.
For production of light vector mesons
the relativistic
corrections play an important role especially at large
$Q^{2}\gg m_{V}^{2}$, and lead to
$Q^{2}$ dependence of $A_{L,T}$ coming from
the large-size asymmetric $q\bar{q}$ configurations:
$A_{L}(\rho^0;Q^{2}=0)\approx 6.5,~
A_{L}(\rho^0;Q^2 = 100\,{\rm GeV}^2) \approx 10,~
A_{T}(\rho^0;Q^{2}=0) \approx 7,~
A_{T}(\rho^0,Q^2 = 100\,{\rm GeV}^2)\approx 12$ \cite{NNZscan}.
Due to an extra factor $z(1-z)$ in the integrand of
(\ref{eq:5}) in comparison with (\ref{eq:4}),
the contribution from asymmetric $q\bar{q}$- configurations to the longitudinal
meson production is considerably smaller.

The integrands in 
Eqs.~(\ref{eq:4}) and
(\ref{eq:5}) contain the dipole cross section,
$\sigma(\xi,r,{\bf{q}})$.
As was mentioned,
due to a very slow onset of the pure perturbative region
(see Eq.~(\ref{eq:1})),
one can easily anticipate 
a contribution to the production amplitude
coming
from the semiperturbative and nonperturbative $r\gsim R_{c}$.
Following the simplest assumption about an additive property
of the perturbative and nonperturbative mechanism of interaction,
we can represent the contribution of the bare pomeron exchange
to $\sigma(\xi,r,{\bf{q}})$ as a sum
of the perturbative and nonperturbative component\footnote{
additive property of a such decomposition of the dipole
cross section has been detaily discussed in \cite{NNPZ97,NNPZZ98}} 
\beq
\sigma(\xi,r,{\bf{q}}) = 
\sigma_{pt}(\xi,r,{\bf{q}})+\sigma_{npt}(\xi,r,{\bf{q}})\,,
\label{eq:8}
\eeq
with the parameterization of both components at small ${\bf{q}}$
\beq
\sigma_{pt,npt}(\xi,r,{\bf{q}})=\sigma_{pt,npt}(\xi,r,{\bf{q}}=0)
\exp\Bigl(-\frac{1}{2}
B_{pt,npt}(\xi,r){\bf{q^{2}}}\Bigr)\,.
\label{eq:9}
\eeq
Here $\sigma_{pt,npt}(\xi,r,{\bf{q}}=0)
= \sigma_{pt,npt}(\xi,r)$ represent the contribution
of the perturbative and nonperturbative mechanisms to the
$q\bar{q}$-nucleon interaction cross section, 
respectively, $B_{pt,npt}(\xi,r)$ are 
corresponding
diffraction slopes.

A small real part of  production amplitudes can be taken
in the form \cite{GribMig}
%
%
\beq
{\rm Re}{\cal M}(\xi,r) =\frac{\pi}{2}\cdot\frac{\partial}
{\partial\xi} {\rm Im}{\cal M}(\xi,r)\,.
\label{eq:9a}
\eeq
%
%

and can be easily included in the production amplitudes
(\ref{eq:4}),(\ref{eq:5})
using substitution
%
%
\beq
\sigma(x_{eff},r,{\bf q})\rightarrow
\biggl (1-i\frac{\pi}{2}\frac{\partial}{\partial~log~x_{eff}}
\biggr)
\sigma(x_{eff},r) = \biggl [1-i\alpha_{V}(x_{eff},r)
\biggr  ]\sigma(x_{eff},r,{\bf q})
\label{eq:10}
\eeq
%
%

The formalism for calculation of $\sigma_{pt}(\xi,r)$
in the leading-log $s$ approximation was developed
in \cite{NZ91,NZ94,NZZ94}. 
The nonperturbative contribution, $\sigma_{npt}(\xi,r)$, 
to the dipole cross section was used in
Refs.~\cite{NZHera,NNZscan,NNPZ97,NNPZZ98} where
we assume that this soft nonperturbative component
of the pomeron is a simple Regge pole with
the intercept, $\Delta_{npt}=0$.
The particular form together with 
assumption of
the energy independent
$\sigma_{npt}(\xi=\xi_{0},r)=\sigma_{npt}(r)$
($\xi_{0}$ corresponds to boundary condition for the gBFKL
evolution, $\xi_{0}=log(1/x_{0})$, $x_{0} = 0.03$)
allows one to successfully describe \cite{NZHera} the
proton structure function at very small $Q^{2}$,
the real photoabsorption \cite{NNZscan} and
diffractive real and virtual photoproduction of light
\cite{NNPZ97} and heavy \cite{NNPZZ98} vector mesons.
A larger contribution of the nonperturbative pomeron
exchange to $\sigma_{tot}(\gamma p)$ vs.
$\sigma_{tot}(\gamma^{*} p)$ can, for example, explain
a much slower
rise with energy
of the real photoabsorption cross section, 
$\sigma_{tot}(\gamma p)$, in comparison
with $F_{2}(x,Q^{2})\propto
\sigma_{tot}(\gamma^{*} p)$ observed at HERA \cite{H1sf,ZEUSsf}.
Besides, the reasonable form of this soft cross section, $\sigma_{npt}(r)$,
was confirmed in the process of the first determination of the dipole
cross section from the data on vector meson
electroproduction \cite{NNPZdipole}. The so extracted dipole cross section
is in a good agreement with the dipole cross section obtained
from the gBFKL dynamics \cite{NNZscan,NZHera}.
Thus, this nonperturbative component of the pomeron
exchange plays a dominant
role at low NMC energies
in the production of the light vector mesons, where the
scanning radius, $r_{S}$ (\ref{eq:1}), is large.
However, the perturbative component of the pomeron become
more important with the rise of energy also in the nonperturbative
region of the dipole sizes.

If one starts with the familiar impact-parameter representation for amplitude of
elastic scattering of the color dipole
%
%
\beq
{\rm Im} {\cal M}(\xi,r,\vec{q})=2\int d^{2}\vec{b}\,
\exp(-i\vec{q}\vec{b})\Gamma(\xi,\vec{r},\vec{b})\,,
\label{eq:11}
\eeq
%
%
then the diffraction slope $B=\left.-
2{d \log {\rm Im}{\cal M}/ dq^{2}}\right|_{q=0}$
%
equals
%
%
\beq
B(\xi,r)= {1\over 2}\langle \vec{b}\,^{2}\rangle =
\lambda(\xi,r)/\sigma(\xi,r)\,,
\label{eq:12}
\eeq
%
%
where
%
%
\beq
\lambda(\xi,r)=\int d^2\vec{b}~
\vec{b}\,^2~\Gamma(\xi,\vec{r,}\vec{b})\, .
\label{eq:13}
\eeq
%
The generalization of the color dipole factorization
formula (\ref{eq:2}) to the diffraction slope of the
reaction $\gamma^{*}p\rightarrow Vp$ reads:
%
%
\beq
B(\gamma^{*}\rightarrow V,\xi,Q^{2})
{\rm Im} {\cal M}(\gamma^{*}\rightarrow V,\xi,Q^{2},\vec{q}=0)=
\int\limits_{0}^{1} dz\int d^{2}\vec{r}\lambda(\xi,r)
\Psi_{V}^{*}(r,z)\Psi_{\gamma^{*}}(r,z)\,.
\label{eq:14}
\eeq
%
%

The diffraction cone in the color dipole gBFKL approach
for production of vector mesons has been detaily studied
in \cite{NNPZZ98}. Here we only present the salient
feature of the color diffraction slope reflecting
the presence of the geometrical contribution from beam
dipole - $r^{2}/8$
and the contribution from the target proton size - $R_{N}^{2}/3$:
%
%
\beq
B(\xi,r)=
\frac{1}{8}r^{2}+\frac{1}{3}R_{N}^{2}+
2\alpha_{\Pom}'(\xi-\xi_{0}) + {\cal O}(R_{c}^{2})\, ,
\label{eq:15}
\eeq
%
%
where $R_{N}$ is the radius of the proton.
For electroproduction of light vector mesons the
scanning radius is larger than the correlation one 
$r\gsim R_{c}$ even for $Q^{2}\lsim 50$\,GeV$^{2}$
and one recovers a sort of
additive quark model, in which the uncorrelated gluonic clouds
build up around the beam and target quarks and antiquarks and
the term $2\alpha_{\Pom}'(\xi-\xi_{0})$
describe the familiar Regge growth of diffraction slope for
the quark-quark scattering.
The geometrical contribution to the diffraction
slope from the target proton size, ${1\over 3}R_{N}^{2}$,
persists for all the dipole sizes,
$r\gsim R_{c}$ and $r\lsim R_{c}$. The last term in (\ref{eq:15}) 
is also associated with the proton size and is negligibly small.

The soft pomeron and diffractive scattering of large color dipole has been
detaily studied in the paper \cite{NNPZZ98}.
Here we assume the conventional Regge rise of the diffraction
slope for the soft pomeron \cite{NNPZZ98},
%
%
\beq
B_{npt}(\xi,r)=\Delta B_{d}(r)+\Delta B_{N}+
2\alpha_{npt}^{'}(\xi-\xi_{0})\,,
\label{eq:16}
\eeq
%
%
where $\Delta B_{d}(r)$ and $\Delta B_{N}$ stand for the contribution
from the beam dipole and target nucleon size. 
As a guidance we take the experimental
data on the pion-nucleon scattering
\cite{Schiz}, which suggest $\alpha'_{npt}=0.15$\,GeV$^{-2}$. 
In (\ref{eq:16}) the proton size contribution
is
%
%
\beq
\Delta B_{N}={1\over 3}R_{N}^{2}\, ,
\label{eq:17}
\eeq
%
%
and 
the beam dipole contribution has been proposed
to have a form 
%
%
\beq
B_{d}(r) = {r^{2} \over 8}\cdot
{r^{2}+aR_{N}^{2} \over 3r^{2}+aR_{N}^{2}}\,,
\label{eq:18}
\eeq
%
%
where $a$ is a phenomenological parameter, $a\sim 1$.
We take $\Delta B_{N}=4.8\,{\rm GeV}^{-2}$.
Then the pion-nucleon diffraction slope is reproduced with
reasonable values of the parameter $a$ in the formula (\ref{eq:18}):
$a=0.9$ for $\alpha'_{npt}=0.15$\,GeV$^{-2}$ \cite{NNPZZ98}. 

Using the expressions (\ref{eq:4}) and (\ref{eq:5}) for
the $T$ and $L$ production amplitudes in conjunction
with Eqs.~(\ref{eq:8}) and (\ref{eq:9}),
we can calculate the differential cross section of
vector meson electroproduction as a function of $t$.

Following the simple geometrical properties
of the gBFKL diffraction slope, $B(\xi,r)$, (see Eq.~(\ref{eq:15})
and \cite{NZZslope}),
one can express its energy dependence through the energy
dependent effective Regge slope, $\alpha_{eff}'(\xi,r)$
%
%
\beq
B_{pt}(\xi,r) \approx \frac{1}{3}<R_{N}^{2}> + \frac{1}{8}r^{2}
+ 2\alpha_{eff}'(\xi,r)(\xi-\xi_{0}).
\label{eq:19}
\eeq
%
%
The effective Regge slope, $\alpha_{eff}'(\xi,r)$,
varies
with energy differently
at different size of the color dipole
\cite{NZZslope};   
at fixed scanning radius and/or $Q^{2}+m_{V}^{2}$,
it decreases with energy. 
At fixed rapidity $\xi$
and/or $x_{eff}$ (\ref{eq:3}), 
$\alpha_{eff}'(\xi,r)$
rises with $r\lsim 1.5$\,fm.  
At fixed energy, it is a flat function 
of the scanning radius.
At the asymptotically large $\xi$ ($W$),
$\alpha_{eff}'(\xi,r)\rightarrow \alpha_{\Pom}'=0.072$\,GeV$^{-2}$.
At the lower and HERA energies, the subasymptotic 
$\alpha_{eff}'(\xi,r)\sim (0.15-0.20)$\,GeV$^{-2}$ and is very
close to $\alpha_{soft}'$ known from the Regge phenomenology
of soft scattering.
It means, that the gBKFL dynamics predicts a substantial rise
with the energy and dipole size, $r$, of the diffraction slope, $B(\xi,r)$,
in accordance with
the energy and dipole size dependence of the effective
Regge slope, $\alpha_{eff}'(\xi,r)$ and due to a presence of the
geometrical component, $\propto r^{2}$, in (\ref{eq:15}) and
(\ref{eq:16}).

Now we will concentrate on the production of radially excited
$2S$ light vector mesons and will study the differential
cross section $d\sigma/dt$ as a function of $t$.
The most important feature of the production of 
$V'(2S)$ vector mesons is the node effect
- the $Q^{2}$ and energy dependent cancellations
from the soft (large size) and hard (small size)
contributions, i.e. from the region
above and below the node position, $r_{n}$,
to the $V'(2S)$ production amplitude.
The strong $Q^{2}$ dependence of these cancellations comes
from the scanning phenomenon (\ref{eq:1}) when
the scanning radius $r_{S}$ for some value of $Q^{2}$
is close to $r_{n}\sim R_{V}$ ($R_{V}$ is the 
vector meson radius).
The energy dependence of the node effect is due to the 
different energy dependence of the dipole cross section
at small ($r<R_{V}$) and large ($r>R_{V}$) dipole sizes.
The strong node effect in production of radially
excited light vector mesons leading to an anomalous
$Q^{2}$ and energy dependence of the production
cross section was demonstrated in Ref.~\cite{NNPZ97}\footnote{
Manifestations of the node
effect in electroproduction on nuclei were discussed earlier, see
\cite{NNZanom} and \cite{BZNFphi}}
Note, that the predictive power is weak
and is strongly model dependent in the region of $Q^{2}$ and energy
where the node effect becomes exact.

There are several reasons to expect that,
for the production of $2S$ light vector mesons, 
the node effect depends on the
polarization of the virtual photon and of the produced vector
meson \cite{NNPZ97}. 
First, the wave functions of $T$ and $L$ polarized (virtual)
photon are different.
Second, different regions of $z$ contribute to the
${\cal M}_{T}$ and ${\cal M}_{L}$.
Third, different scanning radii
for production of $T$ and $L$ polarized vector mesons
and different energy dependence of $\sigma(\xi,r)$ at
these scanning radii
lead to a different $Q^{2}$ and energy dependence of the
node effect in production of $T$ and $L$ polarized 
$V'(2S)$ vector mesons.
Not so for 
the nonrelativistic limit of heavy
quarkonia where the node effect is very weak and
is approximately polarization independent.
There is a weak polarization dependence of the
node effect which is the most marginal for $\Psi'$
production \cite{NNPZZ98} and
this weak 
node effect still leads to a nonmonotonic $Q^{2}$
dependence of the diffraction slope.
For $\Upsilon'$ production the node effect is negligibly small
and is polarization independent with very high accuracy.

There are two possible scenarios for the node effect which
can occur in the $2S$ production amplitude; the undercompensation
and the overcompensation scenario \cite{NNZanom}.
In the undercompensation case,
the $2S$ production amplitude
$\langle V'(2S)|\sigma(\xi,r)|\gamma^*\rangle$
is dominated by the positive contribution coming from small
dipole sizes, $r\lsim r_{n}$ ($r_{n}$ is the node position),
and the $V(1S)$ and $V'(2S)$ photoproduction
amplitudes have the same sign. 
This scenario corresponds namely to the production
of $2S$ heavy vector mesons, $\Psi'(2S)$ and $\Upsilon'(2S)$.
In the overcompensation case,  
the $2S$ production amplitude
$\langle V'(2S)|\sigma(\xi,r)|\gamma^*\rangle$
is dominated by the negative contribution coming from large
dipole sizes, $r\gsim r_{n}$, 
and the $V(1S)$ and $V'(2S)$ photoproduction
amplitudes have the opposite sign. 

The anomalous properties of the diffraction slope come
from the expression (\ref{eq:14}) and will be presented
elsewhere \cite{prepare99}.
The matrix element on l.h.s of
(\ref{eq:14}) represents the well known production
amplitude $\langle V(V')|\sigma(\xi,r)|\gamma^*\rangle$.
As was mentioned, 
the $1S$ production amplitude
is dominated by contribution from
dipole size corresponding to the scanning radius
$r_{S}\sim 3/\epsilon$ (\ref{eq:1}) with the scale parameter
$A\sim 6$ at $Q^{2}=0$ slightly dependent on $Q^{2}$
\cite{NNZscan}. 
However, 
due to $\propto r^{2}$
behaviour of the slope parameter (see (\ref{eq:15}) and
(\ref{eq:16})),
the integrand of the matrix element on the r.s.h of Eq.~(\ref{eq:14}),
$\langle V(1S)|\sigma(\xi,r)B(\xi,r)|\gamma^*\rangle$,
is $\sim r^{5}\exp(-\epsilon r)$ and is peaked by
$r\sim r_{B}=5/\epsilon=5/3r_{S}$.

The node of the radial wave function of the $2S$ states leads
to peculiarities in $t$ dependence of the differential cross
section.
Following the simple geometrical properties
of the diffraction slope (\ref{eq:15}), (\ref{eq:16}),
because of $\propto r^{2}$ behaviour,
the large size negative contribution to the production
amplitude 
from the region above the node position
corresponds to larger value of the diffraction slope than
small size contribution from the region below the node position.
It means, that
the negative contribution to the $V'(2S)$ production amplitude
coming from the region above the node position, $r\gsim r_{n}$,
has a steeper $t$ dependence than the positive contribution
coming from the small size dipoles, $r\gsim r_{n}$.
It can be expressed in a somewhat demonstrative form
as a $t$- dependent production amplitude:
%
%
\beq
 {\cal M}(t) = \alpha\exp(-\frac{1}{2}B_{1}t)-\beta\exp(-\frac{1}{2}
B_{2}t)
\label{eq:20}\, ,
\eeq
%
%
where $\alpha$ and $\beta$ represent the contribution to the matrix
element from the region below and above the node position, respectively.
In (\ref{eq:20}) $B_{1}$ and $B_{2}$ are the effective diffraction
slopes, which correspond to integration over dipole size $r$ from 0
to the position of the node $r_{n}$ and above the node position.
Thus, the inequality $\alpha > \beta$ corresponds to the undercompensation
whereas $\alpha < \beta$ to the overcompensation regime.
The destructive interference of these two amplitudes 
results in  
a decrease of the effective diffraction slope
for the $V'(2S)$
meson production for small $t$ in contrary with the familiar increase
for the $V(1S)$ meson production.
Such a situation is shown in Fig.~1, where we present the model
predictions for the differential cross section as a function
of $t$ for production of $V(1S)$ and $V'(2S)$ mesons at
different c.m.s. energies $W$ and at $Q^{2}=0$.
Real photoproduction measures the purely transverse
cross section.
As was mentioned in the paper \cite{NNPZ97}, using
our wave functions, at $W\lsim 150$\,GeV for $\rho'(2S)$ production
and at $W\lsim 20$\,GeV for $\phi'(2S)$ production,
the forward production amplitude (\ref{eq:2}) is in undercompensation
regime whereas the matrix element 
$<V'(2S)|\sigma(\xi,r)B(\xi,r)|\gamma>$
on r.h.s. of Eq.~(\ref{eq:14})
is in the overcompensation regime.
As the result we predict the negative valued diffraction
slope at $t=0$ and $Q^{2}=0$ 
For this reason and due to destructive interference of 
two contributions to the production amplitude (\ref{eq:20})
with different $t$ dependencies, 
the differential cross section firstly rises
with $t$, flattens at $t\sim (0.1-0.2)$\,GeV$^{2}$
having a maximum.
At large $t$, the node effect is weak in $t$- dependent
production amplitude because of a steeper $t$ dependence
from the large size dipoles and the differential cross section
falls down following the differential
cross section for $V(1S)$ production.
The position of the maximum can be roughly evaluated 
from (\ref{eq:20}) as follows:
%
%
\beq
t_{max} \sim \frac{1}{B-A}log\Biggl [\frac{\beta^{2}}{\alpha^{2}}
\frac{B^{2}}{A^{2}}\Biggr ]
\, ,
\label{eq:21}
\eeq
%
%
with the supplementary condition
%
%
\beq
\frac{\beta}{\alpha} > \frac{A}{B}
\label{eq:22}
\eeq
%
%
where $A=2B_{1}$ and $B=2B_{2}$, $A<B$.
If the condition (\ref{eq:22}) is not fulfilled the differential
cross section $d\sigma/dt$ for production
of $V'(2S)$ vector mesons has no maximum and has a standard
monotonous $t$- behaviour like for production of $V(1S)$ mesons.

The nonmonotonous $t$- behaviour of the differential cross section
for $\rho'(2S)$ and $\phi'(2S)$ production in the
photoproduction limit is strikingly different from
the familiar decrease with $t$ of the differential cross section 
for the $\rho^{0}(1S)$ and $\phi^{0}(1S)$ real photoproduction.
Here we can not insist on
the precise form of the $t$ dependence of the differential
cross section, the main emphasis is on the likely pattern
of the $t$ dependence coming from the node effect.

At larger energies, $W\gsim 150$\,GeV for the
$\rho'(2S)$ photoproduction and
$W\gsim 30$\,GeV for
$\phi'(2S)$ photoproduction, the node effect
becomes weaker and
we predict
the positive valued diffraction slope at $t=0$
because of positive valued matrix elements
$<V'(2S)|\sigma(\xi,r)|\gamma>$
(\ref{eq:2}) and 
$<V'(2S)|\sigma(\xi,r)B(\xi,r)|\gamma>$ on the r.h.s. of Eq.
(\ref{eq:14}).
For this reason,
the nonmonotonous $t$ dependence of the differential cross
section is changed for the monotonous one, but still the
effective diffraction slope decreases towards small $t$
in contrary to the familiar increase for the
$\rho^{0}(1S)$ and $\rho^{0}(1S)$ photoproduction
(see Fig.~1).

Because of a possible overcompensation scenario
for the longitudinally polarized $\rho'(2S)$ and
$\phi'(2S)$ mesons in the forward direction and 
at small $Q^{2}$ (see Ref.~\cite{NNPZ97}),
we present in Fig.~2 the model predictions for the differential
cross sections as a function of $t$ at different energies $W$
and at fixed $Q^{2}= 0.5$\,GeV$^{2}$ for the production of
$T$, $L$ polarized and polarization unseparated $\rho'(2S)$ and
$\phi'(2S)$ mesons. 
As it was mentioned above, at $Q^{2}= 0.5$\,GeV$^{2}$, the node effect
becomes weaker,
the amplitude for $\rho_{T}'(2S)$ and $\phi_{T}'(2S)$
production at $t=0$ 
is in undercompensation regime and the corresponding
slope parameter $B(V_{T}'(2S))$ is positive valued
because of the positive valued matrix element
$<V'(2S)|\sigma(r,z)B(r,z)|\gamma^{*}>$ on the r.s.h. of Eq.
(\ref{eq:14}).
For this reason, 
we predict the standard decrease of
$d\sigma\biggl (\gamma^{*}\rightarrow V_{T}'(2S)\biggr )/dt$ with $t$ 
(see bottom boxes in Fig.~2).
The above mentioned maximum of $d\sigma/dt$ for the undercompensation
regime is absent due to a weaker node effect and because
the condition (\ref{eq:22}) is not fulfilled.

However, for $Q^{2}\lsim 0.5$\,GeV$^{2}$, 
the amplitude for $\rho_{L}'(2S)$ and $\phi_{L}'(2S)$
production in forward direction, $(t=0)$
(and the matrix element
$<V_{L}'(2S)|\sigma(r,z)B(r,z)|\gamma^{*}>$ as well),
is still in overcompensation regime with the positive
valued diffraction slope $B(V_{L}'(2S))$ at small energies
$W\lsim 20$\,GeV.
It follows in anomalous $t$ dependence of 
$d\sigma\biggl (\gamma^{*}\rightarrow V_{L}'(2S)\biggr )/dt$ 
shown if Fig.~2 (middle boxes).
With the increase of $t$, because of the 
above mentioned 
interference
of two different contributions to the production amplitude
with different $t$ dependencies, 
one encounters the exact cancellation of the large and
small distance contributions.
This fact corresponds to the exact node
effect at some $t\sim t_{min}$.
Thus, the differential cross section firstly falls down 
rapidly with $t$,
have a minimum at $t\sim t_{min}$, following by a rise
when the overcompensation scenario of $t$- dependent
production amplitude is changed for the undercompensation one
and the slope parameter becomes to be negative.
At larger $t$, further pattern of $t$- behaviour is practically
the same as the nonmonotonous
$t$ dependence of
$d\sigma\biggl (\gamma^{*}\rightarrow V_{T}'(2S)\biggr )/dt$    
at $Q^{2}=0$ (see Fig.~1).

The position of the minimum, $t_{min}$, in differential cross section
is model dependent and can be roughly estimated from (\ref{eq:20})
%
%
\beq
t_{min} \sim \frac{1}{B-A}log\Biggl [\frac{\beta^{2}}{\alpha^{2}}
\Biggr ]\, .
\label{eq:23}
\eeq
%
%

With our wave functions we find
$t_{min}\sim 0.03$\,GeV$^{2}$ for $\rho_{L}'(2S)$ production
and 
$t_{min}\sim 0.05$\,GeV$^{2}$ for $\phi_{L}'(2S)$ production
at $Q^{2} = 0.5$\,GeV$^{2}$ and at $W= 5$\,GeV.
However, we can not exclude a possibility that 
this minimum will take a place at larger $t$.
At $Q^{2} < 0.5$\,GeV$^{2}$, $t_{min}$ will be also
located at larger values of $t$.
At higher energy, the position of $t_{min}$ is shifted
to a smaller value of $t$ unless the exact
node effect is reached at $t=0$.
At still larger energy, when longitudinally polarized
$2S$ production amplitude is in undercompensation regime,
this minimum disappears and
we predict the pattern of $t$- behaviour of 
$L$ differential cross section very similar to one 
like nonmonotonous $t$ dependence of
$d\sigma\biggl (\gamma\rightarrow V_{T}'(2S)\biggr )/dt$    
in the photoproduction limit described in Fig.~1.
These predicted anomalies can be tested at HERA measuring
the diffractive electroproduction of $2S$ radially excited  
light vector mesons in the separate polarizations, $T$ and $L$. \\




{\large \bf Conclusions}\\

We study the 
diffractive photo- and electroproduction
of ground state $V(1S)$ and radially excited $V'(2S)$ 
vector mesons within the color dipole gBFKL dynamics
with the main emphasis related 
to the differential cross section
$d\sigma/dt$, which is connected with the diffraction slope.
There are two main consequences
of vector meson production
coming from the gBFKL dynamics.
First, the energy dependence of the $1S$ vector meson production
is controlled by the energy dependence of the dipole cross
section which is steeper for smaller dipole sizes.
The energy dependence of the diffraction slope for $V(1S)$
production is given by the effective Regge slope with a small
variation with energy.
Second
the $Q^{2}$ dependence of the $1S$ vector meson production is
controlled by the shrinkage of the transverse size of the virtual
photon and the small dipole size dependence of the color dipole cross
section.
The $Q^{2}$ behaviour of the diffraction slope is given by
the simple geometrical properties, $\sim r^{2}$, coming
from the color dipole gBFKL phenomenology
of the slope parameter.

The diffraction slope for the production of $2S$
light vector mesons shows very interesting and
anomalous behaviour as function of c.m.s. energy $W$
and $Q^{2}$ and will be detaily analysed elsewhere
\cite{prepare99}.
As a consequence of the node in $2S$ radial wave
function, we predict a strikingly different $t$ dependence
of the differential cross section for production of
$V'(2S)$ vs. $V(1S)$ mesons.
The origin is in
destructive interference of the
large distance negative contribution to the
production amplitude from the region
above the node position with a steeper
$t$- dependence and
small distance positive contribution to the
production amplitude from the region
below the node position with a weaker
$t$- dependence.
As a result, at $Q^{2}=0$
(when the $T$ polarized $V_{T}'(2S)$ mesons are only
produced)
as a consequence of the undercompensation scenario for
$T$ polarized forward production amplitude,
we predict
a nonmonotonous $t$- dependence of $d\sigma\biggl 
(\gamma\rightarrow V_{T}'(2S)\biggr )/dt$
and a decreasing effective diffraction slope for $V_{T}'(2S)$ mesons
towards to negative values
at small $t$ in contrary with the familiar increase
for the $V(1S)$ mesons.
The differential cross section 
$d\sigma\biggl (\gamma\rightarrow V_{T}'(2S)\biggr )/dt$ firstly rises with $t$
having a maximum at $t\sim t_{max}$ given
by Eq.(\ref{eq:21}). At large $t$ when the node effect is
weaker $d\sigma(\gamma\rightarrow V_{T}'(2S))/dt$ has the standard
monotonous $t$- behaviour like for production of $V(1S)$
vector mesons.
The position of the maximum is model
dependent and is shifted to smaller values of $t$ with
rising energy and $Q^{2}$ due to a weaker node effect.

For production of $L$ polarized $V_{L}'(2S)$ mesons,
there is overcompensation at $t=0$ leading to
an exact cancellation of the positive contribution from
large size dipoles and the
negative contribution from small size dipoles to the production
amplitude and to
a minimum of the differential cross
section at some value of $t\sim t_{min}$.
The position of $t_{min}$ is given by Eq. (\ref{eq:23}), is
energy dependent and leads to a complicated
anomalous $t$ dependence of
$d\sigma\biggl (\gamma^{*}\rightarrow V_{L}'(2S)\biggr )/dt$
at fixed $Q^{2}$.
Thus, $d\sigma\biggl (\gamma^{*}\rightarrow V_{L}'(2S)\biggr 
)/dt$ firstly falls down with $t$
having a minimum at $t\sim t_{min}$ when the overcompensation scenario
is changed for the undercompensation one. The following pattern
of $t$- behaviour is then the same like for 
$d\sigma(\gamma\rightarrow V_{T}'(2S))/dt$ at $Q^{2}=0$.
These anomalies are also energy and $Q^{2}$- dependent and
can be studied at HERA.

The experimental
investigation of $t$- dependent differential cross section
for real photoproduction ($Q^{2}=0$) of $V'(2S)$ mesons
at fixed target and HERA experiments,
offers an unique possibility to make a choice
between the undercompensation and overcompensation scenarios.
The presence of the minimum in
$t$- dependent
$d\sigma(\gamma\rightarrow V'(2S))/dt$
in a broad energy region from small to
large energies,
corresponds to the overcompensation scenario, whereas its absence
corresponds to the undercompensation scenario.

The position of the node in the radial $(2S)$ wave function can
be tested also by the vector meson data with separate polarizations 
$(L)$ and $(T)$ at $Q^{2} > 0$. 
The existence of the minimum in $t$- dependent
differential cross section is connected again with the 
overcompensation scenario in $(2S)$ production amplitude
whereas the undercompensation scenario reflects the maximum
of $d\sigma/dt$ and/or the standard monotonous $t$- behaviour.

\pagebreak
{\bf Figure captions:}
\begin{itemize}

\item[Fig.~1]
~- The color dipole model
predictions for the differential cross sections
$d\sigma(\gamma^* \rightarrow V(V'))/dt$
for the real photoproduction ($Q^2=0$)
of the $\rho^{0}, \rho'(2S), \phi^{0}$ and $\phi'(2S)$
at different values of the c.m.s. energy
$W$.

\item[Fig.~2]
~- The color dipole model
predictions for the differential cross sections
$d\sigma_{L,T}(\gamma^* \rightarrow V')/dt$ for
transversely (T)
(top boxes) and longitudinally (L)
(middle boxes) polarized radially excited
$\rho'(2S)$, $\phi'(2S)$
and for the polarization-unseparated
$d\sigma(\gamma^* \rightarrow V')/dt=
d\sigma_{T}(\gamma^* \rightarrow V')/dt+\epsilon
d\sigma_{L}(\gamma^* \rightarrow V')/dt$ for
$\epsilon = 1$ (bottom boxes)
at $Q^{2}=0.5$\,GeV$^{2}$ and different values of the c.m.s. energy
$W$.

\end{itemize}


\begin{thebibliography}{99}


\bibitem{KZ91} 
B.Z.Kopeliovich and B.G.Zakharov, {\sl Phys. Rev.} {\bf D44}
(1991) 3466.

\bibitem{KNNZ93} 
B.Z.Kopeliovich, J.Nemchik, N.N.Nikolaev and B.G.Zakharov,
{\sl Phys. Lett.} {\bf B309} (1993) 179.

\bibitem{NNZanom} 
J.Nemchik, N.N.Nikolaev and B.G.Zakharov,
{\sl Phys. Lett.} {\bf B339} (1994) 194.

\bibitem{NNPZ97} 
J.Nemchik, N.N.Nikolaev, E.Predazzi and B.G.Zakharov,
{\sl Z. Phys} {\bf C75} (1997) 71.

\bibitem{NNPZZ98} 
J.Nemchik, N.N.Nikolaev, E.Predazzi, B.G.Zakharov and V.R.Zoller,
{\sl JETP} {\bf 86} (1998) 1054.

\bibitem{NZZslope} 
N.N.Nikolaev, B.G.Zakharov and V.R.Zoller,
{\sl Phys. Lett.}
{\bf B366} (1996) 337

\bibitem{NZZspectrum} 
N.N.Nikolaev, B.G.Zakharov and V.R.Zoller,
{\sl JETP Lett.} {\bf 60} (1994) 694.

\bibitem{DL} 
A.Donnachie and P.V.Landshoff, {\sl Phys. Lett.} {\bf B185}
(1987) 403; \\
J.R.Cuddell, {\sl Nucl. Phys.} {\bf B336} (1990) 1.

\bibitem{Ryskin} 
M.G.Ryskin, {\sl Z. Phys.} {\bf C57} (1993) 89.

\bibitem{KNNZ94} 
B.Z.Kopeliovich, J.Nemchik, N.N.Nikolaev and B.G.Zakharov,
{\sl Phys. Lett.} {\bf B324} (1994) 469.

\bibitem{NNZscan} 
J.Nemchik, N.N.Nikolaev and B.G.Zakharov,
{\sl Phys. Lett.} {\bf B341} (1994) 228.

\bibitem{Brodsky} 
S.J.Brodsky et al., {\sl Phys. Rev.} {\bf D50} (1994) 3134.

\bibitem{Forshaw} 
J.R.Forshaw and M.G.Ryskin, {\sl Z. Phys.} {\bf C} (1995)
to be published.

\bibitem{GLM} 
E.Gotsman, E.M.Levin and U.Maor, 
{\sl Nucl. Phys.}
{\bf B464} (1996) 251.

\bibitem{Kuraev} 
E.A.Kuraev, L.N.Lipatov and S.V.Fadin,
{\sl Sov. Phys. JETP} {\bf 44} (1976) 443; {\bf 45} (1977) 199.

\bibitem{Balitsky} 
Yu.Yu.Balitsky and L.N.Lipatov,
{\sl Sov. J. Nucl. Phys.} {\bf 28} (1978) 822.

\bibitem{Lipatov} 
L.N.Lipatov,
{\sl Sov. Phys. JETP} {\bf 63} (1986) 904; \\ 
L.N.Lipatov, in:{\sl Perturbative Quantum Chromodynamics},
ed. by A.H.Mueller, World Scientific (1989).

\bibitem{NZ94} 
N.Nikolaev and B.G.Zakharov, {\sl JETP} {\bf 78} (1994) 598;
{\sl Z. Phys.} {\bf C64} (1994)631.

\bibitem{NZZ94} 
N.N.Nikolaev, B.G.Zakharov and V.R.Zoller,
{\sl JETP Letters}  {\bf 59} (1994) 6;
{\sl JETP } {\bf 78}  (1994) 866; {\sl Phys. Lett.} {\bf B328}
(1994) 486.

\bibitem{NNN92} 
N.N.Nikolaev, {\sl Comments on Nucl. Part. Phys.} {\bf 21}
(1992) 41.

\bibitem{prepare99} 
J.Nemchik and N.N.Nikolaev,
paper in preparation.

\bibitem{NNPZdipole} 
J.Nemchik, N.N.Nikolaev, E.Predazzi and B.G.Zakharov,
{\sl Phys. Lett.} {\bf B374} (1996) 199.

\bibitem{NZHera} 
N.N.Nikolaev and B.G.Zakharov,
{\sl Phys. Lett. }{\bf B327} (1994) 149.

\bibitem{NZ91} 
N.N.~Nikolaev and B.G.~Zakharov, {\it Z. Phys.} {\bf C49} (1991) 607;
{\it Z. Phys.} {\bf C53} (1992) 331.

\bibitem{Barone} 
V.Barone, M.Genovese, N.N.Nikolaev, E.Predazzi and B.G.Zakharov,
{\sl Z. Phys.} {\bf C58} (1993) 541;
{\sl Int. J. Mod. Phys A}, {\bf 8} (1993) 2779.

\bibitem{NZglue} 
N.N.Nikolaev and B.G.Zakharov,
{\sl Phys. Lett.} {\bf B332} (1994) 184.

\bibitem{GribMig} 
V.N.Gribov and A.A.Migdal,
{\sl Sov. J. Nucl. Phys.} {\bf 8} (1969) 703.

\bibitem{H1sf} 
H1 Collab., I.Abt et al., {\sl Nucl. Phys.}
{\bf B407} (1993) 515.

\bibitem{ZEUSsf} 
ZEUS Collab., M.Derrick et al., {\sl Phys. Lett.}
{\bf } (1993).

\bibitem{Schiz} 
A.Schiz et al.,
{\sl Phys. Rev.} {\bf D24} (1981) 26.\\
J.P.Burq et al.,
{\sl Phys. Lett.} {\bf B109} (1982) 111.

\bibitem{BZNFphi} 
O.Benhar, B.G.Zakharov, N.N.Nikolaev et al.,
{\sl Phys. Rev. Lett.} {\bf 74} (1995) 3565; \\
O.Benhar, S.Fantoni, N.N.Nikolaev et al.,
{\sl Zh. Exp. Teor. Fiz.} {\bf 111} (1997) 769.



\end{thebibliography}
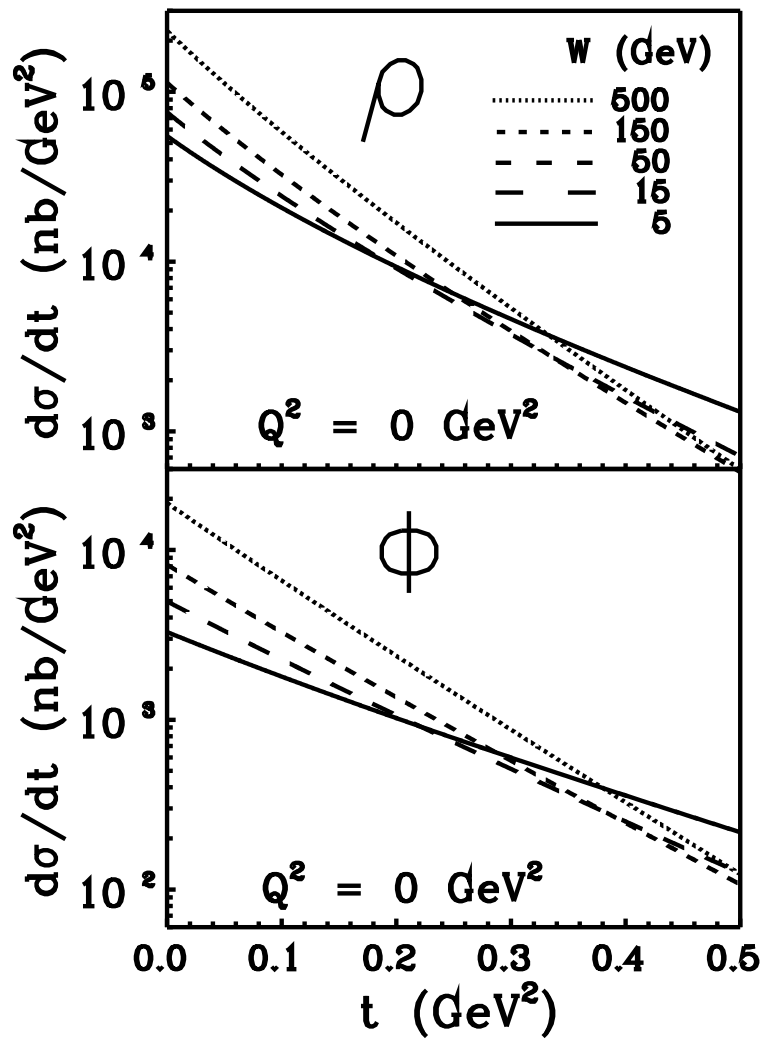
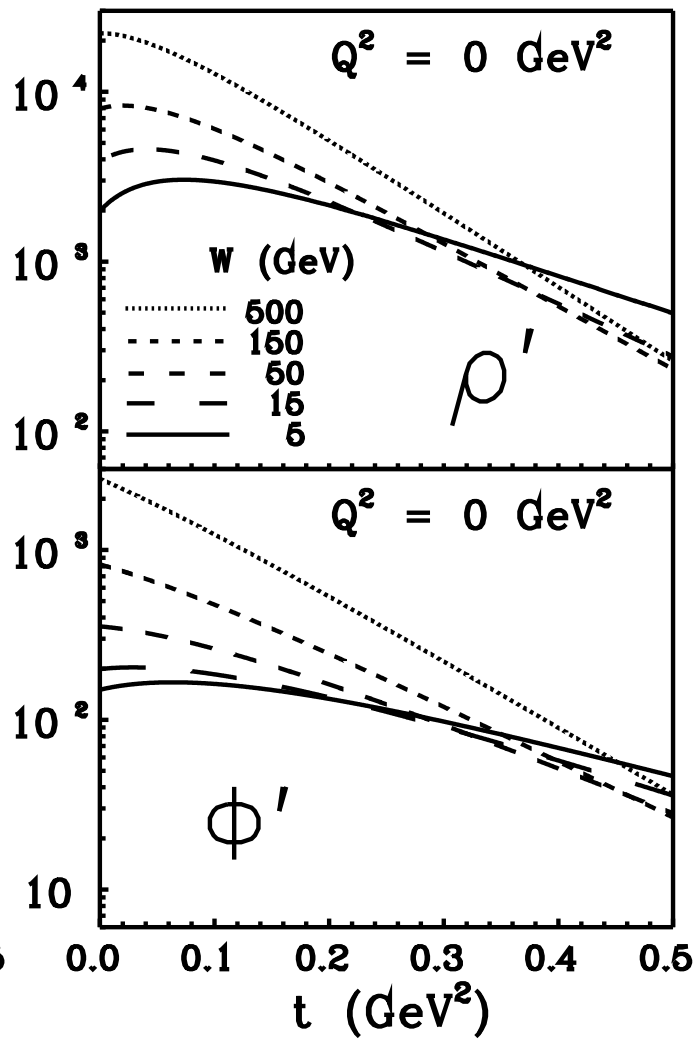
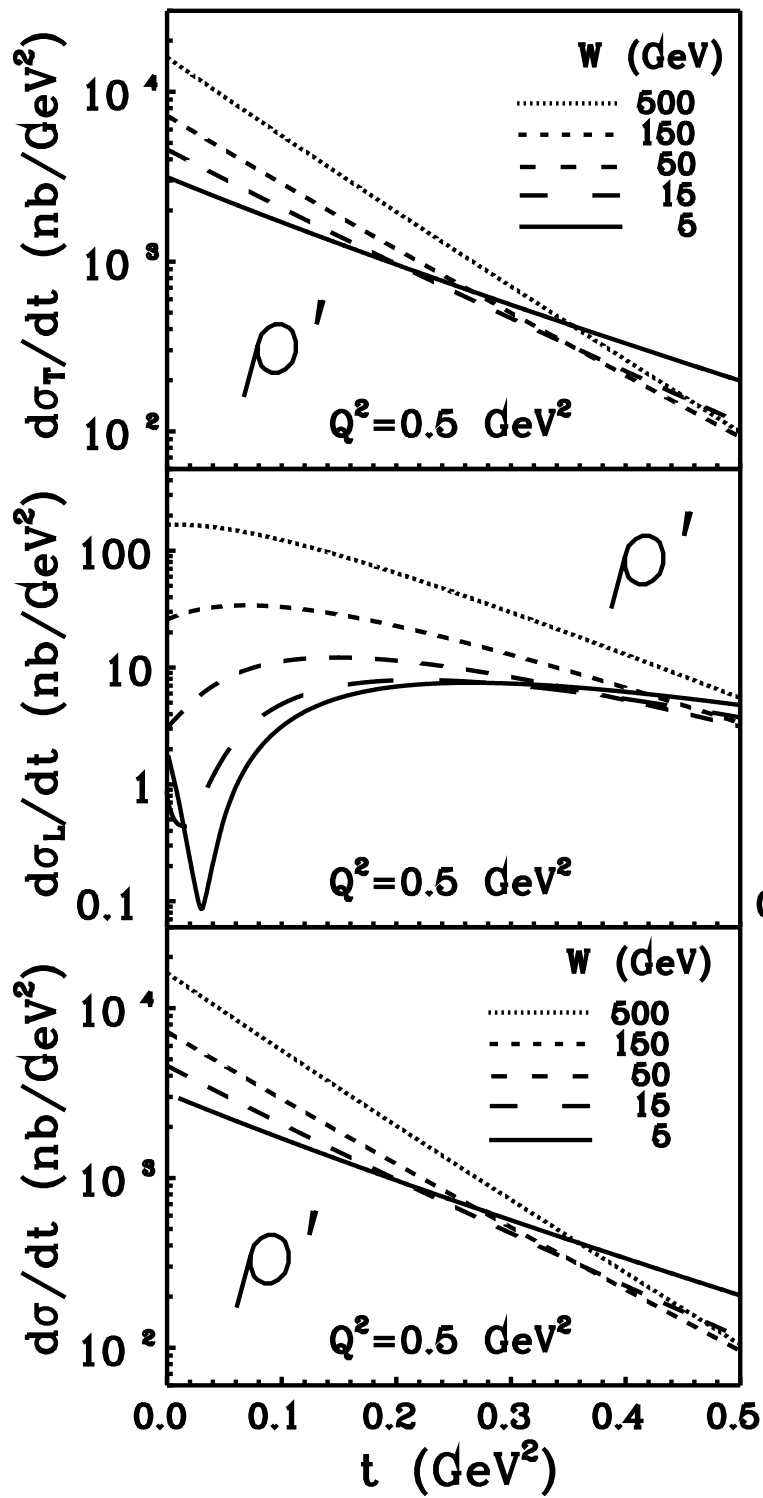
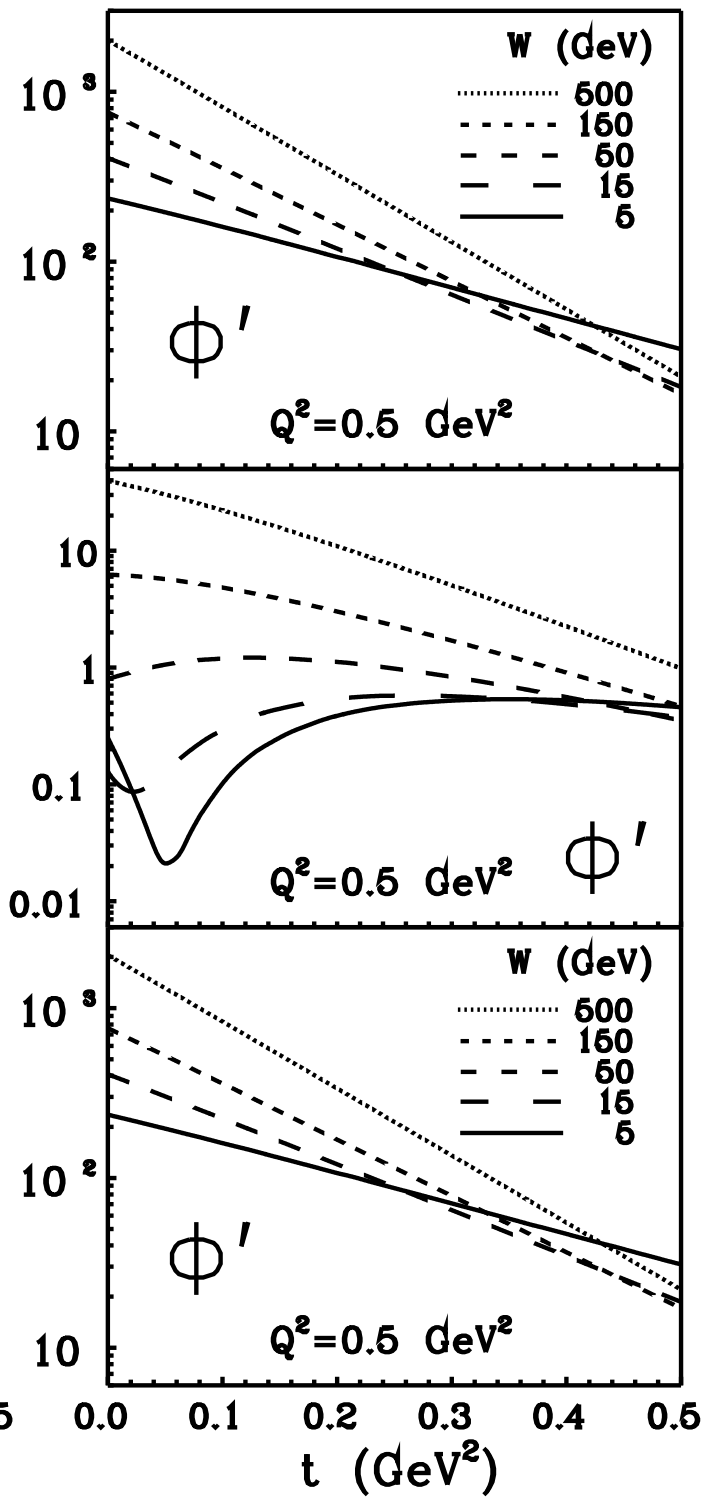
\end{document}